\documentclass{ws-procs961x669}            
\begin{document}
\title{BISOU: a balloon project to measure the CMB spectral distortions}

\author{B.~Maffei$^{*1}$, M.~H.~Abitbol$^2$, N. Aghanim$^1$, J. Aumont$^3$, E. Battistelli$^4$, J. Chluba$^5$, X.~Coulon$^1$, P.~De Bernardis$^4$, M.~Douspis$^1$, J.~Grain$^1$, S.~Gervasoni$^1$, J.~C.~Hill$^{6,7}$, A.~Kogut$^8$, S. Masi$^4$, T.~Matsumura$^9$, C.~O'Sullivan$^{10}$, L.~Pagano$^{11}$, G.~Pisano$^4$, M.~Remazeilles$^5$, A.~Ritacco$^1$, A.~Rotti$^5$, V.~Sauvage$^1$, G.~Savini$^{12}$, S.~L.~Stever$^{13,9}$,  A.~Tartari$^{14}$, L.~Thiele$^{15}$, N.~Trappe$^{10}$}

\address{$^1$Institut d'Astrophysique Spatiale, CNRS-Universit\'e Paris-Saclay, Orsay, 91405, France\\
$^*$E-mail: Bruno.Maffei@universite-paris-saclay.fr}

\address{$^2$University of Oxford, Department of Physics, Denys Wilkinson Building, Oxford OX1 4LS, UK}

\address{$^3$IRAP - CNRS, Toulouse, France}

\address{$^4$Dipartimento di Fisica, Universit\`a di Roma "La Sapienza", Italy}

\address{$^5$JBCA, School of Physics and Astronomy, The University of Manchester, UK}

\address{$^6$Department of Physics, Columbia University, New York, NY, USA 10027}

\address{$^7$Center for Computational Astrophysics, Flatiron Institute, New York, NY, USA 10010}

\address{$^8$NASA - Goddard Space Flight Center, Greenbelt MD 20771 USA}

\address{$^9$Kavli IPMU (WPI), UTIAS, The University of Tokyo, Kashiwa, Chiba, 277-8583 Japan}

\address{$^{10}$Department of Experimental Physics, National University of Ireland, Maynooth, Ireland}

\address{$^{11}$Dipartimento di Fisica e Scienze della Terra - Universit\`a degli Studi di Ferrara, Italy}

\address{$^{12}$Physics and Astronomy Department, University College London, UK}

\address{$^{13}$Okayama University, Kita-ku, Okayama, 700-8530 Japan}

\address{$^{14}$Dipartimento di Fisica "E. Fermi" - Universit\`a di Pisa - INFN, Pisa, Italy}

\address{$^{15}$Department of Physics, Princeton University, Princeton, NJ, 08544 USA}

\begin{abstract}
The BISOU (Balloon Interferometer for Spectral Observations of the Universe) project aims to study the viability and prospects of a balloon-borne spectrometer, pathfinder of a future space mission dedicated to the measurements of the CMB spectral distortions. We present here a preliminary concept based on previous space mission proposals, together with some sensitivity calculation results for the observation goals, showing that a 5-$\sigma$ measurement of the y-distortions is achievable.
\end{abstract}


\bodymatter

\section{Introduction}\label{sec:intro}
With the success of the ESA Planck mission, the concordance cosmological model is established as the reference framework. However, outstanding questions about this model are still unanswered. In particular the simplest inflationary model proposed as the origin of the initial matter perturbations is favoured by Planck measurement of the spectral index and low non-Gaussianity. Nevertheless, it still needs to be confirmed through the measurement of its smoking gun signature: the relic background of primordial gravitational waves. The latter can only be observed through the Cosmic Microwave Background (CMB) polarisation: namely B-modes. The CMB frequency spectrum is another key observable to probe the cosmological model. Its intensity was precisely measured by COBE/FIRAS almost three decades ago \cite{1990ApJ...354L..37M}$^,$\cite{1996ApJ...473..576F}, with deviations limited to $\Delta I/I \approx 10^{-5}$. Since then, not much progress has been achieved in measuring its deviation from a true blackbody. However, while the space mission proposals PIXIE\cite{2011JCAP...07..025K} and PRISTINE (to ESA F-mission call) have not been successful, following two white papers \cite{2021ExA...tmp...42C}$^,$\cite{2021ExA...tmp...76D}, the ESA Voyage 2050 programme has selected this topic amongst its three themes. 

The BISOU (Balloon Interferometer for Spectral Observations of the Universe) project aims to study the viability and prospects of a balloon-borne spectrometer, pathfinder of a future space mission dedicated to the absolute measurement of the CMB spectrum. While PIXIE and PRISTINE were targeting both the measurement of the CMB polarisation, namely a first detection of the CMB B-mode polarisation, and the absolute measurement of the CMB spectral distortions, BISOU's main goal is to perform a first measurement of the later. However, secondary science will also include measurement of the Cosmic Infrared Background (CIB) emission, and potentially the polarisation of the dust up to 2~THz together with Galactic emission lines such as CI, CII, NII and OI.

Taking into account the specificity of a balloon flight in term of requirements and conditions (i.e. residual atmosphere, observation strategy for instance), this CNES Phase~0 study will evaluate if such a spectrometer is sensitive enough to measure at least the Compton y-distortion while consolidating the instrument concept and improving the readiness of some of its key sub-systems.

\section{Science goals}\label{sec:goals}
While the CMB has a nearly perfect blackbody emission spectrum, deviations from it, referred to as spectral distortions, are expected. These distortions encode information about the full thermal history of the Universe from the early stages (primordial distortions from inflation and cosmological recombination lines) until today (star formation and galaxy clusters). Many of these processes are part of our standard cosmological model and are detailed in several publications\cite{2021ExA...tmp...42C}$^,$\cite{2017MNRAS.471.1126A}.

Spectral distortions result from processes that affect the thermal equilibrium between matter and radiation. One of the standard distortions, known as the Compton y-distortion, is created in the regime of inefficient energy transfer (optically thin scattering) between electrons and photons, relevant at redshifts $z<5\times 10^4$. Processes creating this type of distortion are dominated by the inverse-Compton scattering of CMB photons off hot electrons during the epoch of reionization and structure formation, also known as the thermal Sunyaev-Zeldovich (tSZ) effect.

Chemical potential or $\mu$-type distortions, on the other hand, are generated by energy release at earlier stages ($z>5\times 10^4$), when interactions are still extremely efficient (optically thick scattering) and able to establish kinetic equilibrium between electrons and photons under repeated Compton scattering and photon emission processes.

The COBE-FIRAS limit, $|\text{y}| < 1.5\times 10^{-5}$ (95\% C.L.), is roughly one order of magnitude larger than the expected signal\cite{2015PhRvL.115z1301H}, $\text{y}\approx 2\times 10^{-6}$. Signal from $\mu$-type distortions will be even fainter. It is therefore crucial to have reliable models of all the emissions that will be much stronger than these signals in order to properly subtract them.

\subsection{Modeling the signals}

\begin{figure}
\begin{center}
\includegraphics[width=4in]{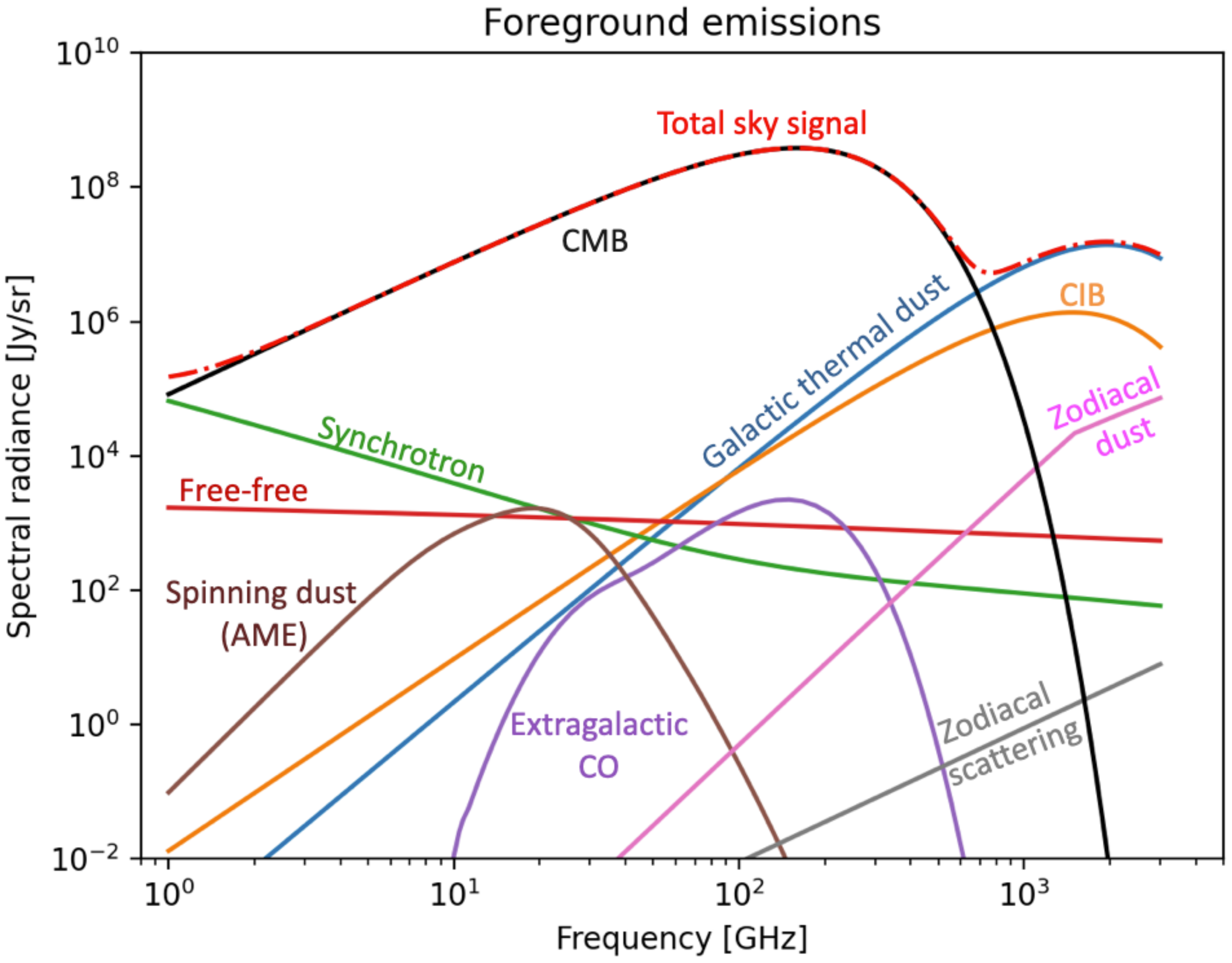}
\end{center}
\caption{Foreground emission contributions with respect to the CMB blackbody emission, together with the sum of all astrophysical signals (CMB + foregrounds).}
\label{aba:fig1}
\end{figure}

Several astrophysical foregrounds contribute meaningfully to the sky signal at frequencies relevant to CMB spectral distortions. In addition to the ones already modeled\cite{2017MNRAS.471.1126A} from Planck data\cite{2016A&A...594A..10P}, additional components from the zodiacal dust have been included. Figure~\ref{aba:fig1} shows the individual contributions together with the sum of all sky signals, including the CMB. At low frequencies (below 70~GHz), the brightest foregrounds are from the synchrotron, the free-free and the so-called anomalous microwave emissions. High frequencies foregrounds (above 100~GHz) are mainly due to the emission of the Galactic thermal dust, the cumulative redshifted emission from thermal dust in distant galaxies, called the cosmic infrared background (CIB), and the zodiacal thermal dust emission. Additional foregrounds contributing to the total sky signal are the cumulative CO emission from distant galaxies at intermediate frequencies and the zodiacal scattering at high frequencies. 

\subsection{CMB spectral distortions modeling}

\begin{figure}
\begin{center}
\includegraphics[width=4in]{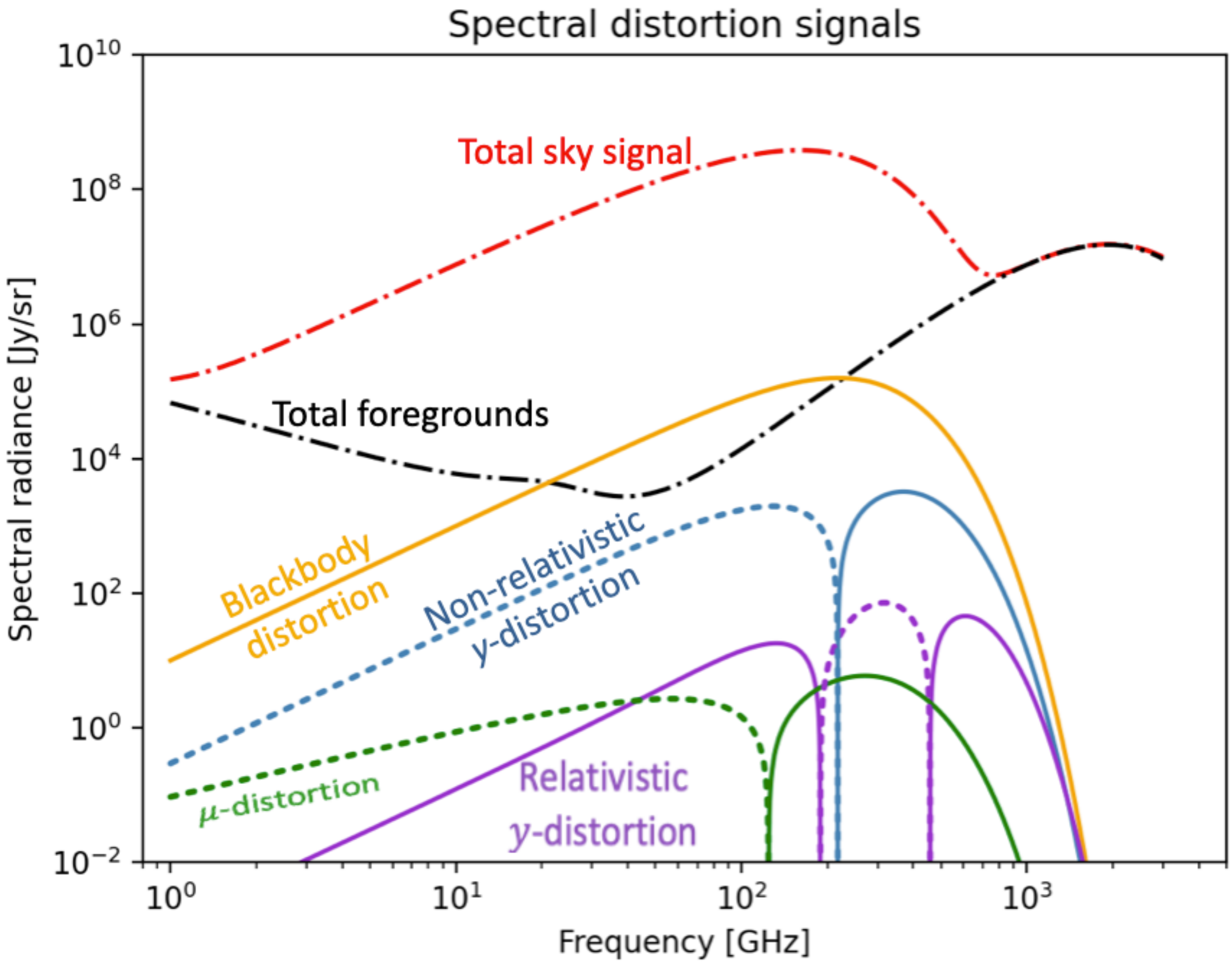}
\end{center}
\caption{Models of the spectral distortions: Black body distortion $\Delta T_{CMB}$ (orange), non-relativistic y-distortion (blue), relativistic y-distortion (purple) and $\mu$-distortion (green). This is compared to the total foreground signal (dashed black) and the total sky signal - Foregrounds + CMB (dashed red).}
\label{aba:fig2}
\end{figure}

Following the same assumptions and processes as the ones presented in Abitbol et al.\cite{2017MNRAS.471.1126A}, figure~\ref{aba:fig2} shows the signals associated with the various distortions of the spectrum, with respect to the total emission of the foregrounds and the total sky signal (including the CMB emission). 

Four contributions are considered and modeled. First a blackbody distortion that represents a first order temperature deviation $\Delta T_{CMB}$ to the true CMB blackbody spectrum. Then a cumulative thermal SZ y-distortion, including both standard non-relativistic and relativistic contributions (from Hill et al.\cite{2015PhRvL.115z1301H}). The intracluster medium (ICM), the intergalactic medium and reionization contributions are included in the Compton-y signal. A relativistic correction to thermal SZ distortion\cite{2015PhRvL.115z1301H}$^,$\cite{2017MNRAS.471.1126A} is modeled using the moment-based approach. Finally the chemical potential $\mu$-distortion is generated assuming only signals from acoustic damping and adiabatic cooling.
The $r$-type distortion, sometimes called ``residual'' distortion, is expected to have a contribution smaller than the $\mu$-type distortion and is not represented in figure~\ref{aba:fig2}.

\section{Instrument concept}\label{sec:inst}

\begin{figure}
\begin{center}
\includegraphics[width=2.4in]{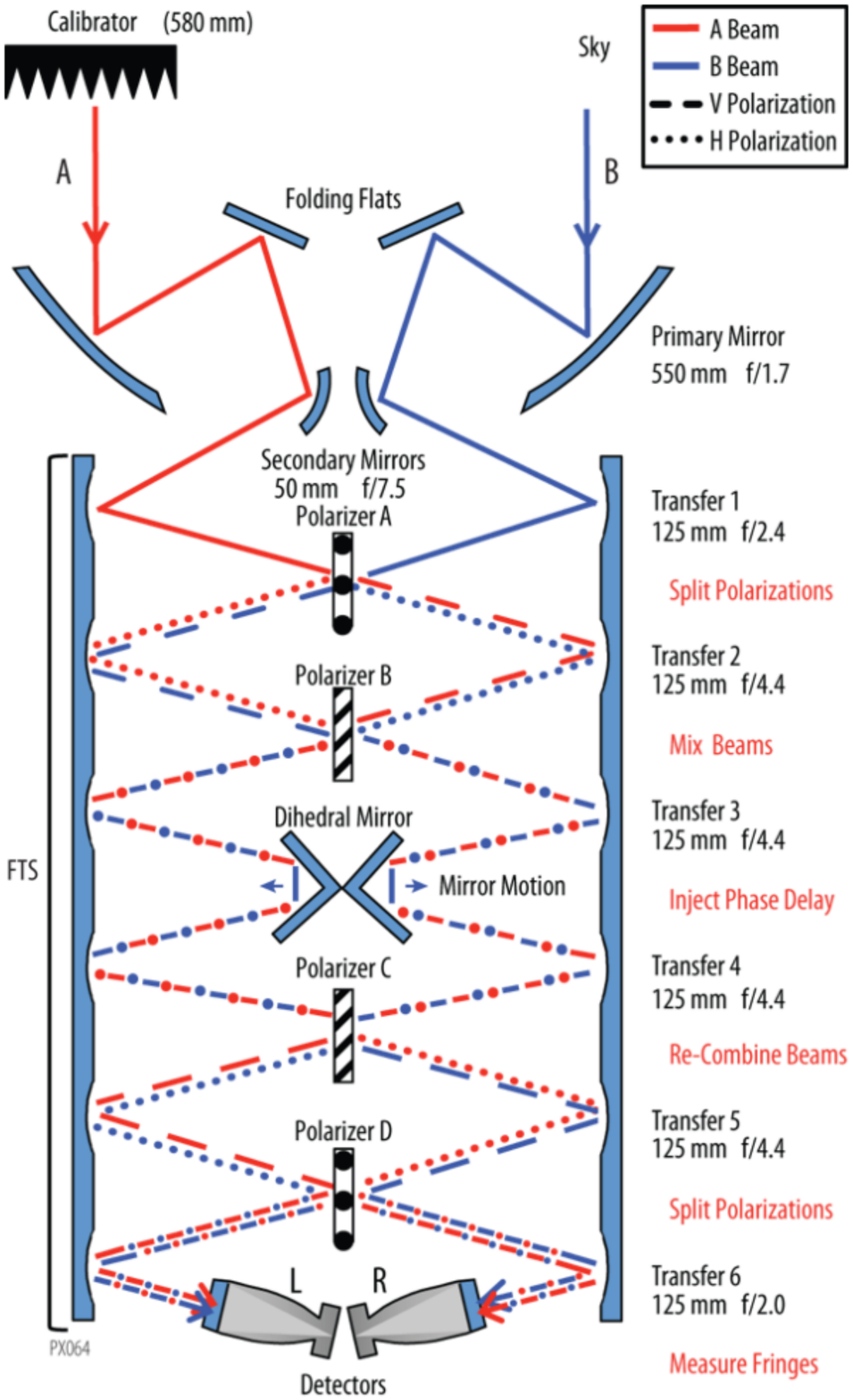}
\includegraphics[width=2.4in]{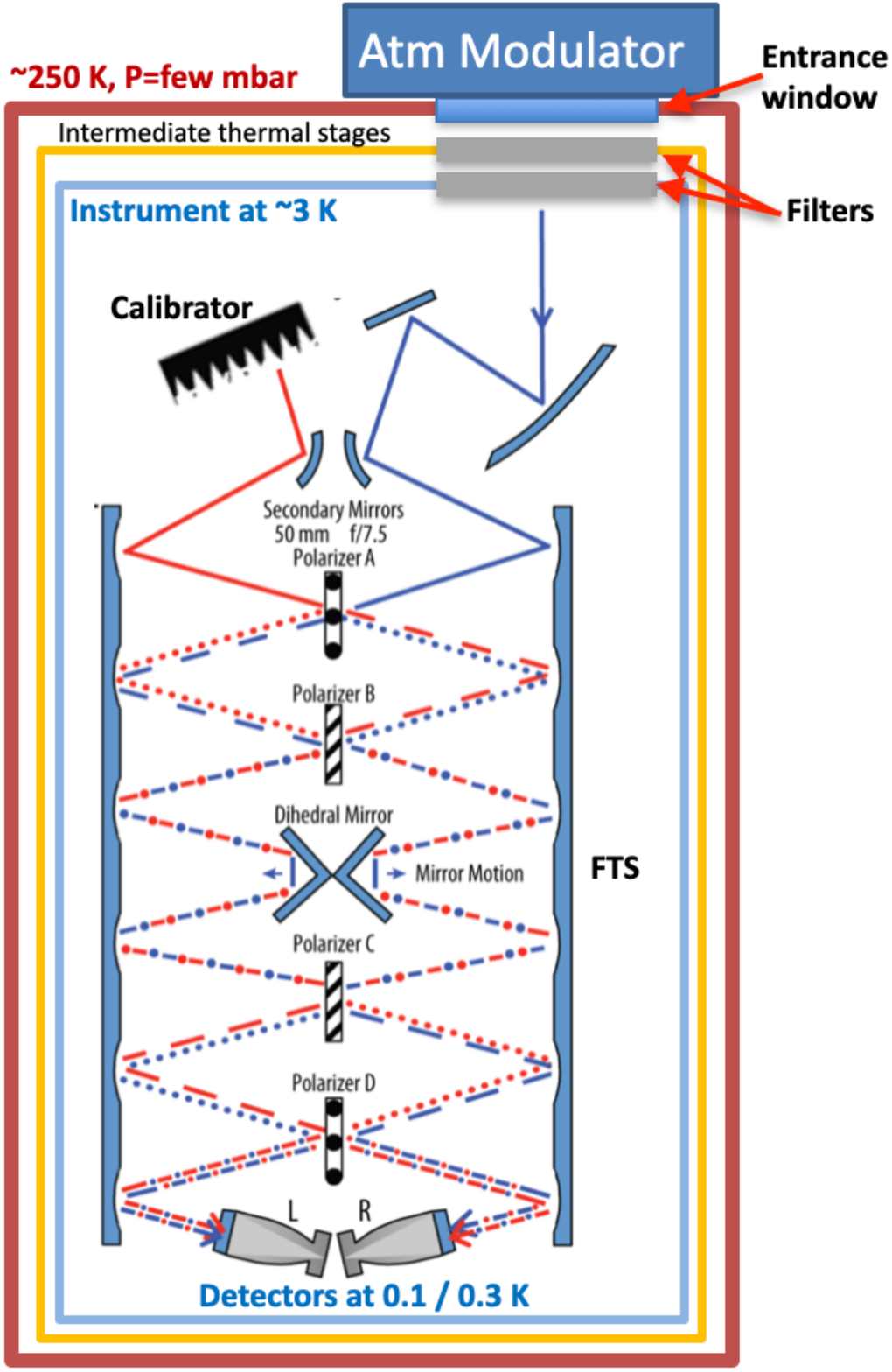}
\end{center}
\caption{Left: Original instrument concept from the PIXIE space mission proposal \cite{2011JCAP...07..025K}. Right: BISOU instrument scheme.}
\label{aba:pixie}
\end{figure}

The initial concept is based on a similar one that had been proposed by the PIXIE team \cite{2011JCAP...07..025K} (shown in figure~\ref{aba:pixie} left) and also used for the PRISTINE F-class ESA mission proposal. The instrument is a Fourier Transform Spectrometer (FTS) with two inputs and two outputs. Both inputs are going through a separate telescope, both sets of optics being identical in order to minimise the systematics. Because PIXIE has several observation modes, an external calibrator (cooled blackbody) could be located in front of one of the apertures at any one time. In the ``spectral distortion'' mode, for which the absolute spectrum needs to be measured, one input is directed towards the sky, the second towards a blackbody whose temperature is set to the CMB one, 2.726~K, so that a differential measurement can be achieved. This process can be alternated in order to cancel any asymmetry in both optical systems. After going through a set of polarisers, each of the two outputs is focused on a dual-polarisation multimoded bolometric detector, so that each of the four detectors measures an interference fringe pattern between orthogonal
linear polarizations from the two input beams. In order to limit the photon noise, the whole instrument is cooled to about 3~K, the detectors being at sub-K temperature. 

\subsection{Balloon specificity}\label{sec:ball}
The previous studies for the instrument concept were performed on the basis of a space mission. For the case of a balloon project, the conditions, the requirements and therefore the instrument concept will be different. Stratospheric balloon projects might be considered to be in near-space conditions, but there are still many differences, some of these being the potential flight time, access to the sky, thermal environment, etc... 

A few will have strong impacts on the payload concept. First the residual atmosphere at an altitude of about 40~km will be of the order of 3~mbar with a temperature of the order of 270~K. This will not only lead to an additional photon noise contribution, but also to the necessity of hosting the overall instrument inside a dewar in order to keep it at a temperature of the order of 3~K under vacuum. Even with a small pressure difference, the dewar will then need to have a window through which the telescope will point towards the sky (see figure~\ref{aba:pixie} right). The outside dewar shell being at ambient temperature, intermediate thermal stages will be necessary with thermal filters in order to limit the thermal background load and potential straylight. 

This is also preventing the calibrator to be outside the dewar and will then have to be located inside in order to keep it at a very steady temperature of about 3~K.

Due to the mass, power and dimensions constraints of a typical balloon payload, it is unlikely that two sets of optics can be used with a telescope primary diameter of 35 to 40~cm allowing for a multi-moded beam FWHM of about 1.5~deg.

\subsection{Gondola, dewar and cooling chain}
Assuming that the balloon flight will be provided by CNES, our study is based on the use of the CARMEN gondola\cite{HEMERA} design that has been used lately for the PILOT balloon project \cite{2019ExA....48..265M} for instance, allowing for a maximum all-included payload mass of 750~kg (including ballast). While CNES has so far flown these types of payload for a typical 35-hour flight duration, it is planned that a first 5-day test-flight will happen in 2022. We will therefore base our sensitivity calculations on the assumption that this type of flight will be available in the future. 

The whole instrument needs to be cooled to about 3~K, where the detectors only will be cooled to a sub-K temperature. A trade-off is being made in order to define the optimum one, but it is highly likely that 300~mK will be low enough (section~\ref{sec:phot}) for the focal plane, allowing for the use of a $^3$He sorption  cooler. Due to mass and power limitations, mechanical coolers for balloon platforms are not yet mature enough, even if some developments are being investigated, most notably by NASA. We chose a proven cooling solution using liquid helium to cool the overall instrument, between 4 and 2~K depending on the bath pressure, using the natural low pressure at high altitude. Several intermediate thermal stages and shields cooled with the helium vapour retrieved from the bath will allow for heat load reduction on the 3~K stage. Typical intermediate temperature stages are 150~K, 80~K and 20~K.

\subsection{Atmosphere}
As already mentioned, the residual atmosphere at 40~km altitude will still create a large photon noise contribution with respect to the extremely faint signal that we are trying to observe. For instance, atmospheric effects at ground level and even at balloon altitudes are shown in Masi et al.\cite{2021cosmo} with respect to spectral distortion signals. The signal level due to the atmosphere is still 2 to 3 orders of magnitude higher that the non-relativistic y-distortion signal depending on the frequency. In theory, assuming a reliable model, this contribution could be removed.

More important, will be the variations of the atmosphere with altitude, observation elevation and time. This cannot be modelled and will therefore need an atmosphere modulation strategy at a high enough frequency to be compatible with the variation timescale. Such a modulator is already used for COSMO\cite{2021cosmo}, but for BISOU, another modulator, such as a cold internal beam stirrer is also being considered. 

For the time being, at that stage of the study, the atmosphere is not yet considered in our sensitivity and systematic effects calculations.

\subsection{The instrument}

Following the conditions of a balloon project (section~\ref{sec:ball}), several points had to be adapted and trade-offs have already been made. With emphasis on the spectral distortions, these measurements will always be performed against the calibrator. Due to the fact that we are using only one telescope to save mass (with therefore an unbalanced optical system), the calibrator will be kept fixed in front of one of the FTS inputs for simplicity and risk mitigation. 

With the help of an instrument model being developed, and assuming some initial parameters such as telescope diameter, flight duration and observation efficiency, or again spectral resolution, calculations presented in section~\ref{sec:sens} are showing that the sensitivity is strongly dependant on three key parameters: the temperature and emissivity of the window, and the maximum frequency of observation as the photon noise is increasing with frequency. The implications of these issues will be discussed in the following sections. 
However, we see straight away that the dewar window, the warmest optical element, needs to have a very low emissivity, a temperature as low as possible and be as thin as possible to minimise the systematics.

In order to use a very thin window, a valve at the entrance of the dewar, opening only when the pressure difference is small (at a certain altitude), could be used. Such a technique was used for the Archeops\cite{2002APh....17..101B} balloon project for instance. 

The emissivity of the window could be reduced by the application of a copper/gold grid as was done on the ArT\'eMiS\cite{2006SPIE.6275E..03T} ground based camera.

Finally while the dewar will be at a temperature ranging from 300 to 270~K, the window could have a mount thermally isolated from the structure and actively cooled with the vapour coming from the helium bath. Such a development will be studied with a dedicated R\&D programme. This should bring the window temperature below 200~K.

\section{Preliminary sensitivity estimates}\label{sec:sens}

\begin{figure}
\begin{center}
\includegraphics[width=5in]{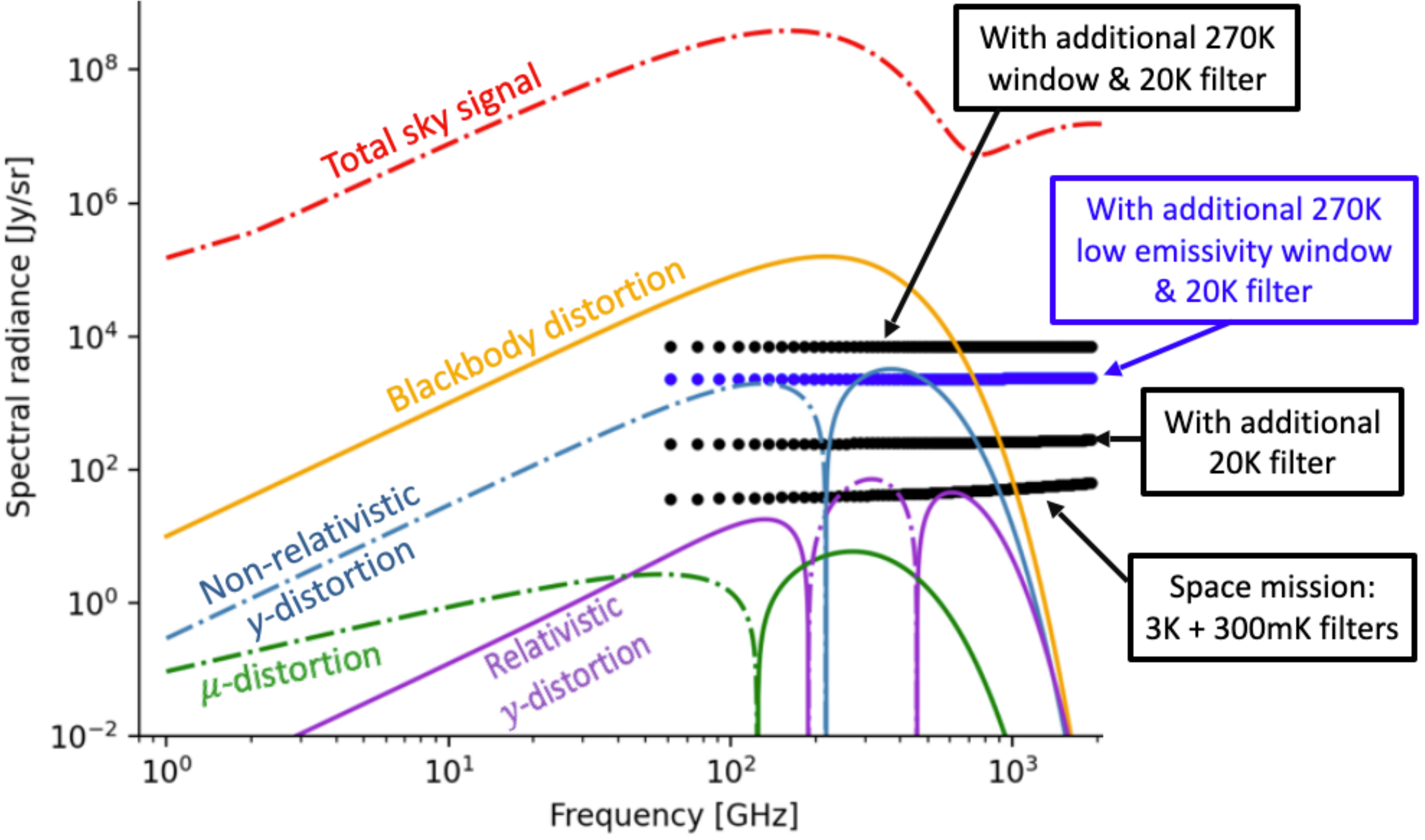}
\end{center}
\caption{Estimated instrument sensitivity for different concepts, compared to various spectral distortion signals. All calculations are assuming a 5-day flight. The highest sensitivity assumes a space mission, while the lowest sensitivity is for a balloon flight for which the noise is limited by the emission of the dewar window with pessimistic characteristics.}
\label{aba:fig4}
\end{figure}

\subsection{Photometric model}\label{sec:phot}
A preliminary photometric model has been developed, taking into account only the dewar window and a minimum of two spectral filters on the lowest temperature stages (supposed to be at 20 and 3~K and with a top-hat transmission profile) for the optical input. The emission of all the components in the optical path is modeled as blackbody emission (at the temperature of the component) multiplied by the emissivity of that component at the specified temperature.
The load on the detector is estimated by adding the power contributions of the optical components as well as the contribution from the sky, using for each:

\begin{equation*}
P(\nu,T)=\int_{\nu_{min}}^{\nu_{max}}eff(\nu)A\Omega(\nu)\epsilon(\nu,T) B(\nu,T) d\nu
\end{equation*}

where $A\Omega$ is the throughput of the multi-moded optics, $eff$ the transmission/efficiency  of the optical system, $B$ the blackbody function, and for this first version of our model, the emissivity $\epsilon$ is taken as a constant for each component.

We integrate the power received by the detector over the full range of frequencies. 
In order to assume the worst case scenario for the detector load, when the optical path difference of the Fourier Transform Spectrometer is null, the power is integrated over the full frequency range that we will observe.

The total Noise Equivalent Power (NEP) is then calculated by adding in quadrature the photon NEP from the signal arriving on the detector and the detector NEP which is assumed to be about four times lower than the photon NEP. For the time being, we assume that the other contributions to the total NEP are small in comparison to the photon one. Depending on the assumptions made, for the temperature and emissivity of the optical components for instance, the total NEP is of the order of a few $10^{-16}$~W~Hz$^{-1/2}$ for a space mission to about $10^{-14}$~W~Hz$^{-1/2}$ for a balloon configuration, high enough to allow for the use of detectors at a temperature of about 300~mK.

\subsection{Results}

Following classic calculations\cite{2011JCAP...07..025K}, from the total NEP, the detected noise for a fixed integration time $\tau$ is given by:

\begin{equation*}
\delta P=\frac{NEP_{total}}{\sqrt{\tau/2}}
\end{equation*}

where $\tau$ is taken to be 90~hours, corresponding to a flight duration of 5 days with 75\% observation efficiency. 
From this, the noise at the detector may in turn be referred to the specific intensity, leading to an equivalent instrumental sensitivity that can be compared to spectral emissions. These results are shown on figure~\ref{aba:fig4} for different instrument configurations. For all these configurations, a FTS spectral resolution of 15~GHz, a lowest frequency $\nu_{min}$=60~GHz, and a linear tapered filter to limit the high frequency contribution above 600~GHz have been used.

Figure~\ref{aba:fig4} shows the impact of the window on the sensitivity as it will be the warmest element on the optical chain. Starting from a ``space configuration'' where the warmest element might be around 3~K (filter at 3~K and 300/100~mK with 0.1\% emissivity each), which will be the most sensitive configuration, followed by a configuration with an added filter at T=20~K, then the same configuration with the window at T=270~K added, this with 2 emissivity assumptions for the window, one with $\epsilon$=0.1\% like all other filters, and a lower one with $\epsilon=3\times 10^{-4}$ (in blue), something that we hope to achieve. 

We then conclude that in order to have a first detection of the y-parameter with a decent signal-to-noise ratio, the window will need to be cooled further as previously discussed. 

Another parameter which is also crucial, is the maximum observation frequency $\nu_{max}$. All the photon contributions being higher with increasing frequency, the overall sensitivity will drop with increasing $\nu_{max}$. On the other hand, in order to retrieve the ``easy'' science goals, such as the y-parameter spectral distortions, $\Delta T_{CMB}$ and $T_{CIB}$ for the measurement of the Cosmic Infrared Background, measurements have to go to high frequencies. Therefore a trade-off needs to be reached in order to optimise the S/N for each parameter.

This work has started and will be presented in future publications. However, we can already say that, not taking into account the atmosphere, S/N ratios of 5 for the y-distortions, about 30 for $\Delta T_{CMB}$ and about 10 for $T_{CIB}$, can be reached if $\nu_{max}$ were to be 1200~GHz.

Obviously a more detailed study will need to take place for a better optimisation, including more sophisticated models and better parameter knowledge.

\section{On-going and future work}
Aside from more detailed model of the instrument and sensitivity calculations, several points need to be addressed by the end of the Phase~0 study. 

The next important hurdle to cross is understanding how to deal with the atmosphere. More accurate models and data for high altitude are being gathered within the consortium but as already stated, a modulator seems inevitable. Concept work and the study of the associated effects of this modulator will be paramount.

Other points to be developed (not exhaustive) can be listed as:
\begin{itemize}
    \item Observation plan and scanning strategy;
    \item Calibration strategy, together with the accurate characterisation and control of the calibrator;
    \item Study on how to increase the sensitivity, by increasing the number of detectors for instance, or having two channels;
    \item A first cut to the study of the systematic effects, in order to answer at least if the unbalanced optical system between the two inputs will not be a limiting factor;
    \item Updated science case.
\end{itemize}

\section{Conclusion}
So far, assuming that the atmospheric problem can be solved, it seems that according to our preliminary estimates, such a balloon borne instrument will be able to have valid scientific outputs, and not be limited to a technological demonstrator for future space missions. Namely, a measurement of the CIB, a better constraint on $\Delta T_{CMB}$, and more importantly a first 5-$\sigma$ measurement of the y-distortions if not better.

If these results are confirmed by the end of Phase~0, a proposal to CNES and other funding bodies will be submitted to start a more detailed Phase~A study, for a first test-flight on the horizon 2026. 

\section*{Acknowledgments}
The authors acknowledge the support of the French space agency, Centre National d'Etudes Spatiales (CNES).

Kavli IPMU is supported by World Premier International Research Center Initiative (WPI), MEXT, Japan.
\bibliographystyle{ws-procs961x669}
\bibliography{BISOU-MG16}

\begin{thebibliography}{10}

\bibitem{1990ApJ...354L..37M}
J.~C. {Mather} {\em et~al.}, {A Preliminary Measurement of the Cosmic Microwave
  Background Spectrum by the Cosmic Background Explorer (COBE) Satellite}, {\em
  ApJL} {\bf 354}, p. L37 (May 1990).

\bibitem{1996ApJ...473..576F}
D.~J. {Fixsen}, E.~S. {Cheng}, J.~M. {Gales}, J.~C. {Mather}, R.~A. {Shafer}
  and E.~L. {Wright}, {The Cosmic Microwave Background Spectrum from the Full
  COBE FIRAS Data Set}, {\em ApJ} {\bf 473}, p. 576 (December 1996).

\bibitem{2011JCAP...07..025K}
A.~{Kogut}, D.~J. {Fixsen}, D.~T. {Chuss}, J.~{Dotson}, E.~{Dwek},
  M.~{Halpern}, G.~F. {Hinshaw}, S.~M. {Meyer}, S.~H. {Moseley}, M.~D.
  {Seiffert}, D.~N. {Spergel} and E.~J. {Wollack}, {The Primordial Inflation
  Explorer (PIXIE): a nulling polarimeter for cosmic microwave background
  observations}, {\em JCAP} {\bf 2011}, p. 025 (July 2011).

\bibitem{2021ExA...tmp...42C}
J.~{Chluba}, M.~H. {Abitbol}, N.~{Aghanim} {\em et~al.}, {New horizons in
  cosmology with spectral distortions of the cosmic microwave background}, {\em
  Experimental Astronomy}  (May 2021).

\bibitem{2021ExA...tmp...76D}
J.~{Delabrouille}, M.~H. {Abitbol}, N.~{Aghanim} {\em et~al.}, {Microwave
  spectro-polarimetry of matter and radiation across space and time}, {\em
  Experimental Astronomy}  (July 2021).

\bibitem{2017MNRAS.471.1126A}
M.~H. {Abitbol}, J.~{Chluba}, J.~C. {Hill} and B.~R. {Johnson}, {Prospects for
  measuring cosmic microwave background spectral distortions in the presence of
  foregrounds}, {\em MNRAS} {\bf 471}, 1126 (October 2017).

\bibitem{2015PhRvL.115z1301H}
J.~C. {Hill}, N.~{Battaglia}, J.~{Chluba}, S.~{Ferraro}, E.~{Schaan} and D.~N.
  {Spergel}, {Taking the Universe's Temperature with Spectral Distortions of
  the Cosmic Microwave Background}, {\em Physical Review Letters} {\bf 115}, p.
  261301 (December 2015).

\bibitem{2016A&A...594A..10P}
{Planck Collaboration}, {Planck 2015 results. X. Diffuse component separation:
  Foreground maps}, {\em A\&A} {\bf 594}, p. A10 (September 2016).

\bibitem{HEMERA}
CNES, {\em CNES Balloons capabilities}
  \url{https://www.hemera-h2020.eu/facilities-2/cnes-balloons/}.

\bibitem{2019ExA....48..265M}
A.~{Mangilli} {\em et~al.}, {Inflight performance of the PILOT balloon-borne
  experiment}, {\em Experimental Astronomy} {\bf 48}, 265 (December 2019).

\bibitem{2021cosmo}
S.~{Masi} {\em et~al.}, {The COSmic Monopole Observer (COSMO)}, {\em These
  proceedings}   (2021).

\bibitem{2002APh....17..101B}
A.~{Beno{\^\i}t} {\em et~al.}, {Archeops: a high resolution, large sky coverage
  balloon experiment for mapping cosmic microwave background anisotropies},
  {\em Astroparticle Physics} {\bf 17}, 101 (May 2002).

\bibitem{2006SPIE.6275E..03T}
M.~{Talvard} {\em et~al.}, {ArTeMiS: filled bolometer arrays for next
  generation sub-mm telescopes}, in {\em Society of Photo-Optical
  Instrumentation Engineers (SPIE) Conference Series\/},  eds. J.~{Zmuidzinas},
  W.~S. {Holland}, S.~{Withington} and W.~D. {Duncan}, Society of Photo-Optical
  Instrumentation Engineers (SPIE) Conference Series, Vol.~6275June 2006.

\end{thebibliography}

\end{document}